COVER SHEET

Paper Number: 244

Title: **Investigating the Composite/Metal Interface and its Influence on the Electrical Resistance Measurement**

Authors:  Pedram Gharghabi
               Juhyeong Lee
               Tinsley Colmer
               Michael S. Mazzola
               Thomas E. Lacy
               Joni Kluss

**ABSTRACT**

The advantages introduced by carbon fiber reinforced polymer (CFRP) composites has made them an appropriate choice in many applications and an ideal replacement for conventional materials. The benefits using CFRP composites are due to their lightweight, high stiffness, as well as corrosion resistance. For this reason, there is a fast growing trend in using CFRP composites for aircraft and wind turbine structural applications. The replacement of the conventional aerospace-grade metal alloys (aluminum, titanium, magnesium, etc.) with CFRP composites results in new challenges. For example, an aircraft during flight is prone to be struck by lightning. To withstand the injection of such massive amount of energy, adequate electrical properties, mainly electrical conductivity, is required. In fact, electrical conductance (or its reciprocal, resistance) is a critical parameter representing any material change and it can be considered an index for health monitoring. In this paper, AS4/8552 carbon/epoxy laminated composites were injected with two types of electrical currents, impulse current and direct current. The change in measured electrical resistance was recorded. A significant resistance drop occurred after electrical current injections. Furthermore, four-point flexural tests were performed on these composites to correlate an electrical resistance change with a potential flexural property change. There was no clear trend between a resistance change and flexural strength/modulus change of the test coupons, regardless of current injection. However, it was observed that the injection of the current affects the contact resistance such that its resistance decreases. It is argued, then, that previous reports of electrical resistance change is an artifact of the measurement system rather than an observation of a change in the composite material. The cause for the resistance change is discussed.

__________

Pedram Gharghabi, Department of Electrical and Computer Engineering
Juhyeong Lee, Department of Aerospace Engineering
Michael S. Mazzola, Department of Electrical and Computer Engineering
Thomas E. Lacy, Jr., Department of Aerospace Engineering
Joni Kluss, Department of Electrical and Computer Engineering
Mississippi State University, Mississippi State, MS 39762.

# INTRODUCTION

Composites have been widely used in electrical engineering primarily as an insulator. To be used in electrical applications, composites are required to have acceptable electrical properties, mainly a reasonable electrical conductivity. Due to their excellent mechanical properties, the CFRP composites are gaining a great interest in the construction of the modern aircraft and wind turbines [1]. The replacement of the conventional metallic aircraft structures with CFRP composites introduces concerns regarding their electrical conductivity. For instance, the outer skin of an aircraft should act as a shield against a lightning strike and prevents the penetration of the lightning current, which can lead to catastrophic damages. Lightning-induced damage is mainly thermal damage, and Joule heating is a primary indicative of lightning thermal damage, which is proportional to electrical resistance (thus, inversely proportional to conductivity). A carbon fiber reinforced epoxy (or carbon/epoxy) composite is representative of a CFRP composite presently used in aircrafts. For carbon/epoxy composites, carbon fiber functions as a reinforcement and improves mechanical strength, and the epoxy matrix is a supporting material. While carbon fibers have a relatively good electrical conductivity, the epoxy matrix acts like an insulator and tends to block any current conduction. In addition, carbon/epoxy composites have different electrical properties in each direction, and measuring their resistance is not a straightforward task. A proper contact must be provided to accurately measure the resistance. The most important variable pertinent to an electrical contact is its resistance, and it must be considered as an integral part of the electrical system or circuit.

An electrical contact is meant to provide a continuous flow of current through the interface between two contacting mediums. The mechanisms and processes that take place at the interface to allow passage of electrons across the interface are very complex. The surfaces that are meeting are not ideally flat, and the interface is established through their asperities which form localized contact. The surface roughness is affected by a cluster of separated conducting paths, and their number and sizes of these paths are a function of the force and pressure holding the surfaces together, and the current density [2]. In other words, when the surfaces are contacted, first the larger asperities of the opposing surfaces meet. Figure 1 shows the schematic.

These spots provide the conducting paths for the electrical current to flow. Therefore, the current is forced and restricted to transfer though these spots at the interface of the meeting conductors. The real area of contact is then, determined by the total number of the individual spots conducting the current [3]. As the pressure holding the contacting the surfaces increases, the resistance drops until it stabilizes, and even after the pressure is released the contact resistance at the interface between composite and the electrode is much better than its initial value [4]. The constriction of the current through these spots results in an increase of the electrical resistance, which is referred as the constriction resistance. This resistance is determined by the number and size of each individual spot. Larger number of these spots leads to a lower contact resistance [3]. The most important feature of an electrical contact is the heat generated due to passage of current. The Joule heating, and the associated temperature gradient, is the most plausible factor to explain the detrimental effect of the current on the change of contact characteristics. The temperature rise at the interface is equal or slightly higher than the rest of the conductors in a good

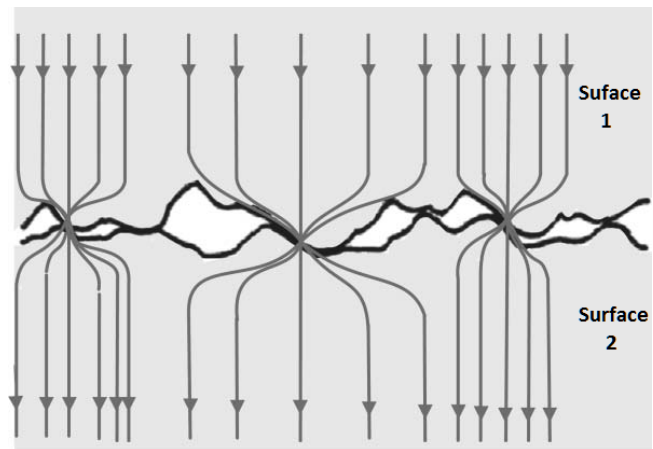
Figure 1: Asperities meeting at the contact interface

connection; whereas, in a poor electrical connection, highly localized hot spots form that increases the bulk temperature. It can lead to thermal damage and deterioration of the contacting interface, causing higher resistances. The heat generated at the interface is determined by the square of current multiplied by the constriction resistance [5]. All things considered, the temperature rise is a function of the dimensions of the contacting surfaces, the current density, and voltage drop across the contact. It must be noted that thermal conductivity decreases as the temperature rises. Therefore, at higher currents, larger thermal energy is generated, but this heat is not dissipated at the same rate [5].

## RESISTANCE MEASUREMENT

Electrical resistance is an inherent property of a composite which is determined by the sample dimensions and its resistivity. The resistance in a CFRP composite is considered a monitoring index and is commonly believed to be correlated with material change [6]–[8]. Different types of damages are attributed to the change of the material resistance caused by flow intensified current: 1) fiber breakage increases the resistance along the fibers, 2) delamination between layers increases through thickness resistance because it separates the contacting fibers of the adjacent layers, and 3) fiber-matrix debonding increases the resistance transverse to the fibers.

However, the electrical resistance measurements of a composite is not a straightforward task, where one can simply place the probes on the electrode and read the resistance. The simplest technique to measure the resistance is the two-probe method. This technique is based on Ohm's law, where the current is injected into one electrode and extracted from another electrode. Then, the voltage difference is measured across two electrodes. This resistance measurement includes the resistance of the composite sample, wires and connections, and the contact resistance. The resistance of the wires and connections are some magnitude of order smaller than that of the composite, and hence they can be ignored. Since the value of the contact resistance is unknown and usually hard to determine, this technique is not accurate, and it is not recommended in many applications.

**Direct Current Test**

The experimental setup comprises a direct current (DC) source, Thermovision Camera, and a voltmeter. The rated values of the DC source are 60 V and 500 A, and the output can be either constant voltage or current. In this experiment, the target applied current was set, and the voltage was regulated automatically to supply the desired current with regards to the change of the resistances. Therefore, based on Ohm's law, any change of the voltage input is an indication of a change of the resistance in the circuit. The ThermoVision A20M is an infrared camera used for temperature measurement of the coupons (test samples) during application of the current. Figure 2 shows the experimental setup for DC current injection test.

The edges of carbon/epoxy composite test coupons were sanded to expose carbon fibers. Two copper plates were placed on the edges of the test coupon. Then, the electrodes, copper plate, and test coupon were clamped to form the terminals (current injection and extraction points). The test was started at 1 A input current, and continued at incremental steps of 1 A until the temperature limit of the system, 400°C, was reached. The current was retained for 60 seconds, and the variation of the voltage supply was recorded. As discussed before, the change of the voltage at constant current is a manifestation of the resistance change; if the resistance drops, the voltage decreases to maintain the constant current, and conversely the increase of the resistance causes the voltage to rise.

**Impulse Current Test**

A customized impulse generator was designed to perform the low impulse current (IC) tests [9]. The generator is capable of producing maximum 300 A peak amplitude within tens of microseconds. This setup is comprised of a charging unit, charging storage module, a switch, and the measurement instrument. Four 100 μF capacitors are responsible for storing the electrical energy, and they are connected in parallel to ensure the maximum output current. A multi-chip transistor can safely switch the current and trigger the discharge. The DC voltage source provides the electrical energy and is connected to the capacitors. The charging voltage of the DC source determines the energy transferred to the capacitors. This energy, when trigged, is discharged in to the test sample, and thus, the charging voltage determines the output impulse current. A wire coil current transformer measures the impulse current and records the waveform the current injected to the test coupon. The schematic of the impulse generated can be seen in Figure 3.

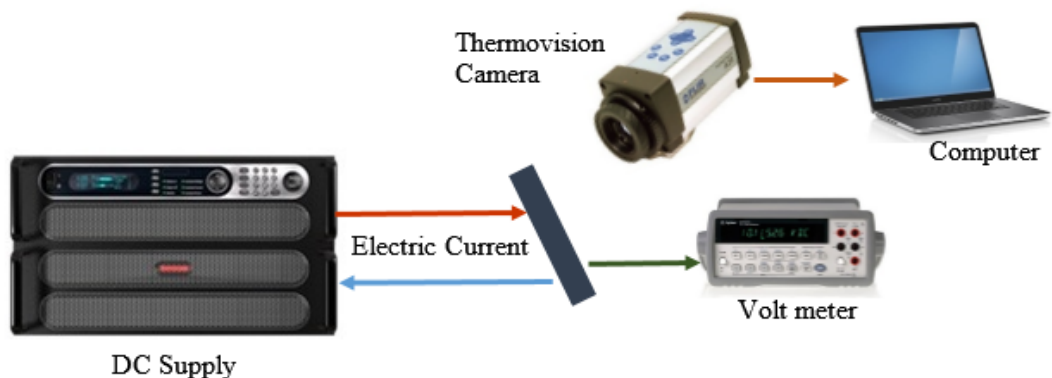

Figure 2. Experimental setup for DC current injection test.

A copper plate was placed between the terminal and the ends of the test coupon to provide a uniform conduction path. Also, a thin layer of epoxy was removed to provide a better electrical contact. The first voltage level of the impulse test started from 50 V and it was repeated three times. The current limit was set to 300 A, and the voltage was increased at 50 V incremental steps until the limit was reached. The current waveform was recorded and the resistance was measured after each shot.

A typical current waveform applied to a unidirectional test coupon is shown in Figure 4. It was recorded at the charging voltage of 300 V, and the temporal characteristics of the applied current can be measured from the recorded waveform. As can be seen, 284 A is the peak amplitude of the current, the time to the peak amplitude (i.e. rise time) is 4.68 µs, and the time to the half value on the tail (i.e. tail time) is 17.6 µs.

**Composite Test Coupon Fabrication**

The nine-ply AS4/8552 carbon/epoxy composites were fabricated based upon the manufacturer's recommended cure procedure (initial cured at 107°C for 1 h and post-cure at 177°C for 2 h; [10]). Two material configurations were considered: a) $[0]_9$ unidirectional and b) $[(90/0)_4/90]$ cross-ply composites.

**RESULTS AND DISCUSSIONS**

The test coupons were subjected to IC and DC, and the resulting resistance change was recorded. For each type of current test, a minimum of four unidirectional and cross ply coupons were injected. The average value of the readings at each step for both unidirectional and crossply coupons are shown in Figure 5.

As can be seen, the cross-ply coupons start from a larger resistance compared to the unidirectional coupons. It is due to the fact that the resistance is primarily determined by the fiber direction of the top layer, and in case of a cross-ply sample they are perpendicular to the preferred path of current. The results are consistent with the previous report [9]. Both unidirectional and cross-ply coupons experienced more than 80% resistance drop, which is a significant electrical change. The DC currents were injected on the unidirectional and cross-ply coupons. Figure 6 shows the average value of the resistances at each current level.

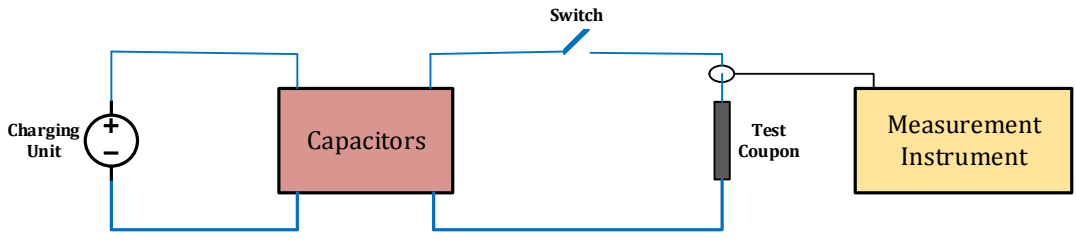

Figure 3. Schematic of impulse current test setup.

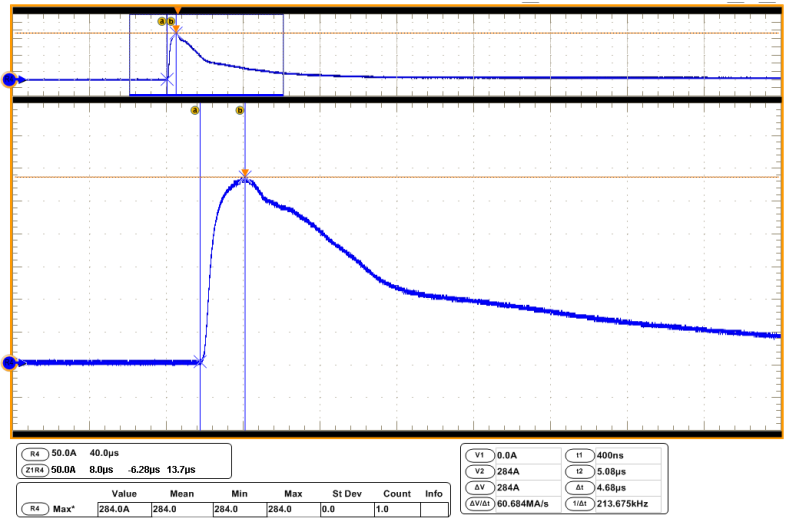

Figure 4. Electric current vs. time, applied to a unidirectional coupon.

Similar to the case where the test coupons were subjected to the IC, the measured resistance decreases sharply to about 20% of its original value. The coupons tested with IC experienced slightly larger resistance change compared to the coupons subjected to DC current. The difference between resistance measurements of the IC versus DC current can be associated with their temporal characteristics. IC rises rapidly to the peak amplitude and attenuates in the range of tens of microseconds. Whereas, DC has much lower amplitude and the current is sustained for longer time period. Malucci [11] has argued that the contact impedance is a function of the frequency of the applied current, which denotes that the conducting path contains resistive and capacitive components. In [12], it was shown that the heat generated due to an IC mostly takes place at the regions very close to the interface, and the area far from the contact region does not experience a considerable temperature change; however, injection of DC current for a longer time duration results in development of thermal stresses across the conductors. This fact implies that the thermal analysis might be ignored when a composite is subjected to low IC. The effect of the time duration of the applied current on the composite materials is also studied in [13], and it was concluded that the electromagnetic effects of the current suppress the thermal effect, when the applied current has a transient nature and is applied in a very short time; however, the thermal effects are dominant when the current is applied for a longer time duration.

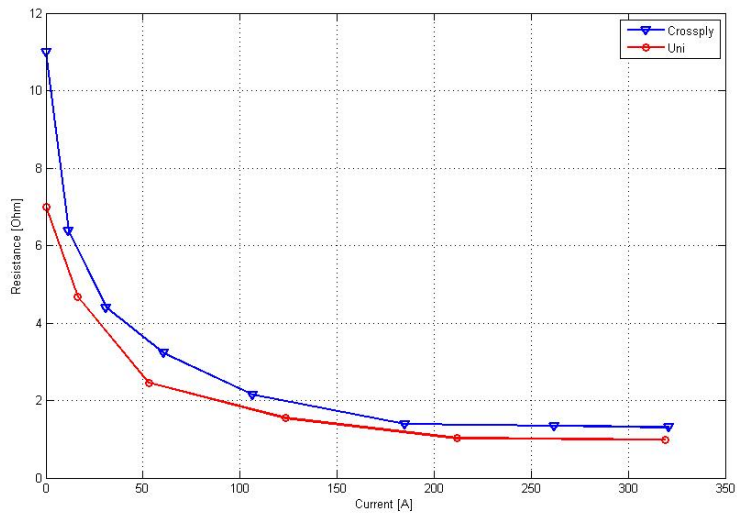

Figure 5: Average value of resistance measurements of IC test unidirectional and cross-ply test coupons.

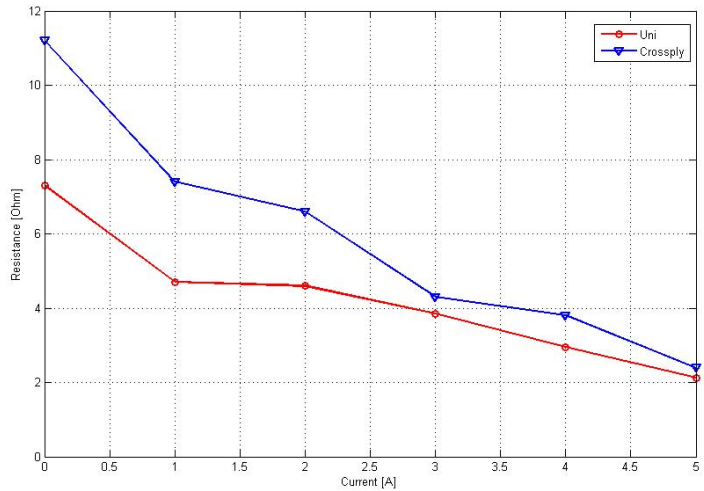

Figure 6: Resistance change against DC current for unidirectional cross-ply coupons.

**Four-point Flexural Test**

A series of four-point flexural tests were performed under displacement control (2 mm/min) in accordance with ASTM standard D6272-02 [14]. The carbon/epoxy coupons with 115×13 ×1.5 mm$^2$ (length × width × thickness) dimensions were tested with a 90 mm support span. Flexural strengths and moduli of the baseline test coupons (prior to current injection) were compared with those from IC-injected and DC-injected coupons. TABLE 1 includes a comparison of flexural strengths and moduli of AS4/8552 carbon/epoxy composites prior to current injection, after the application of IC and DC. Regardless of the current injected, the unidirectional test coupons exhibited flexural strength and modulus nearly two times those of the cross-ply test coupons. A tendency between current injection and flexural strength/modulus was not clear due to small variations (see Figure 7). The composite flexural properties seemed to be somewhat insensitive to current injection. This suggests that an electrical resistance change due to current injection leads to an insignificant change in flexural properties.

TABLE 1. COMPARISON OF FLEXURAL STRENGTHS AND MODULI OF AS4/8552 CARBON/EPOXY COMPOSITES PRIOR TO CURRENT INJECTION AND SUBJECTED TO IC AND DC

| Case | Sample [1] | # of Replicate Test | Current Injection [2] | Avg. Flexural Strength (MPa) | Avg. Flexural Modulus (GPa) | COV [3] Flexural Strength | COV [3] Flexural Modulus |
|---|---|---|---|---|---|---|---|
| 1 | Uni | 6 | No | 1,262 | 121.0 | 6.9% | 5.9% |
| 2 | Uni | 6 | IC | 1,354 | 124.5 | 2.3% | 3.9% |
| 3 | Uni | 6 | DC | 1,366 | 125.0 | 3.8% | 2.5% |
| 4 | Crossply | 4 | No | 697 | 62.2 | 2.6% | 5.7% |
| 5 | Crossply | 6 | IC | 711 | 61.1 | 4.9% | 5.2% |
| 6 | Crossply | 6 | DC | 681 | 59.0 | 3.6% | 5.1% |

[1] Uni (Unidirectional) and Cross (Crossply) test coupons.
[2] No (no current injection), IC (impulse current), and DC (direct current).
[3] Coefficient of variation indicating standard deviation divided by average of a flexural property.

## CONCLUSIONS

A series of AS4/8552 carbon/epoxy laminated composites were subjected to impulse current (IC) and direct current (DC). The resistance change of the carbon/epoxy composite was compared with that prior to current injection. The test coupons subjected to either IC or DC showed significant electrical resistance drops, coupons tested with IC experienced around 90%, and coupons tested with DC experienced around 80%. It was suggested that the change in resistance measurements are due to the change of resistance at the interface between the contact and the coupons. The sudden drop of the measured resistance does not have any reflection related to change of composite material property. The resistance at the contact interface can be improved, but it can never be completely eliminated, and it has to be always considered as an inseparable component of the measurements. To investigate correlation between an electrical resistance change and mechanical property change, four-point flexural tests were conducted. The measured flexural strengths and moduli were insensitive to an electrical resistance change caused by the current injection.

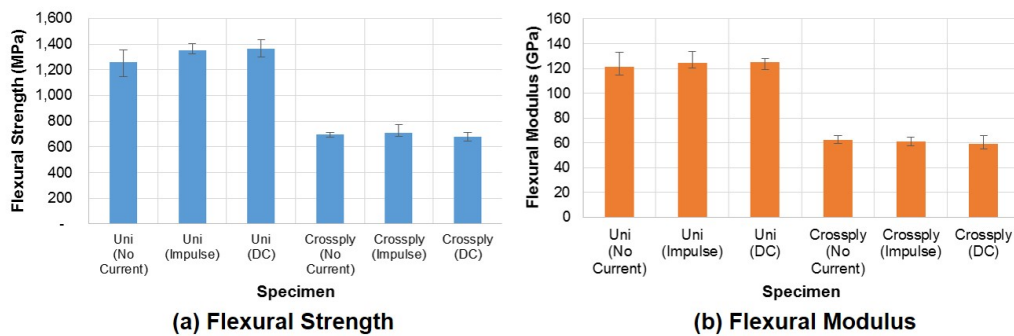

Figure 7. (a) Flexural strengths and (b) moduli of AS4/8552 carbon/epoxy composites, prior to current injection, subjected to impulse current, or DC.

In the current study, quasi-static flexural tests were performed where the load is applied so slowly and consequently the structure also deforms very slowly, so that the inertia force can be ignored. Although neither IC nor DC injection tests showed considerable variations in flexural strengths and moduli, dynamic test may do so. The next step is to perform, cyclic fatigue tests on carbon/epoxy composites consistent with our previous test coupons.